# Hard x-ray standing-wave photoemission study of the interfaces in a $BiFeO_3/La_{0.7}Sr_{0.3}MnO_3$ superlattice

H. P. Martins,[a,b] S. A. Khan,[c] G. Conti,[a,b] A. A. Greer,[a] A. Y. Saw,[a] G. K. Palsson,[d] M. Huijben,[e] K. Kobayashi,[f] S. Ueda,[g,h] C. M. Schneider,[i,a,j] I. M. Vishik,[a] J. Minár,[c] A. X. Gray,[k] C. S. Fadley,[a,j,1] and S. Nemšák*,[b]

[a] *Department of Physics, University of California, Davis, CA 95616, USA*

[b] *Advanced Light Source, Lawrence Berkeley National Laboratory, Berkeley, CA 94720, USA*

[c] *New Technologies-Research Center, University of West Bohemia, 306 14 Pilsen, Czech Republic*

[d] *Department of Physics, Uppsala University, 752 37 Uppsala, Sweden*

[e] *University of Twente, MESA+ Institute for Nanotechnology, 7500 AE Enschede, Netherlands*

[f] *Quantum Beam Science Directorate, Japan Atomic Energy Agency, 1-1-1 Kouto, Sayo, Hyogo 679-5148, Japan*

[g] *NIMS Beamline Station, SPring-8, National Institute for Materials Science, 1-1-1 Kouto, Sayo, Hyogo, 679-5148 Japan*

[h] *Research Center for Functional Materials, NIMS, 1-1 Namiki, Tsukuba, Ibaraki 305-0044, Japan*

[i] *Peter-Grunberg-Institut 6, Forschungszentrum Jülich, 52425 Jülich, Germany*

[j] *Materials Science Division, Lawrence Berkeley National Laboratory, Berkeley, CA 94720, USA*

[k] *Department of Physics, Temple University, Philadelphia, PA 19122, USA*

*e-mail: snemsak@lbl.gov

**Abstract**. Hybrid multiferroics such as $BiFeO_3$ (BFO) and $La_{0.7}Sr_{0.3}MnO_3$ (LSMO) heterostructures are highly interesting functional systems due to their complex electronic and magnetic properties. One of the key parameters influencing the emergent properties is the quality of interfaces, where varying interdiffusion lengths can give rise to different chemistry and distinctive electronic states. Here we report high-resolution depth resolved chemical and electronic investigation of BFO/LSMO superlattice using standing-wave hard X-ray photoemission spectroscopy in the first-order Bragg as well as near-total-reflection geometry. Our results show that the interfaces of BFO on top of LSMO are atomically abrupt, while the LSMO on top of BFO interfaces show an interdiffusion length of around 1.2 unit cells. The two interfaces also exhibit different chemical gradients, with the BFO/LSMO interface being Sr-terminated by a spectroscopically distinctive high binding energy component in Sr 2p core-level spectra, which is spatially contained within 1 unit cell from the interface. From the electronic point of view, unique valence band features were observed for bulk-BFO, bulk-LSMO and their interfaces. Our X-ray optical analysis revealed a unique electronic signature at the BFO/LSMO interface, which we attribute to the coupling between those respective layers. Valence band decomposition based on the Bragg-reflection standing-wave measurement also revealed the band alignment between BFO and LSMO layers. Our work demonstrates that standing-wave hard x-ray photoemission is a reliable non-destructive technique for probing depth-resolved electronic structure of buried layers and interfaces with sub-unit-cell resolution. Equivalent investigations can be successfully applied to a broad class of material such as perovskite complex oxides with emergent interfacial phenomena.

---

[1] Deceased 1st August 2019.

# 1. Introduction

Transition metal oxides are known for their complex electronic structure. The interweaving of orbitals that forms the valence bands gives rise to myriad interesting physical properties [1,2], which have been extensively explored in materials science. The interplay between charge, spin, orbital, and lattice degrees of freedom present in these material systems can be successfully manipulated in artificial complex oxides heterostructures, leading to new physical properties and phenomena [3–5]. The full understanding and eventual control of such properties is desirable, although the path to that can be additionally complex when the interplay is between already rich materials.

$BiFeO_3$ (BFO) is the most well-known room-temperature multiferroic [6,7]. Ever since the electric control of antiferromagnetic domains in $BiFeO_3$ at room temperature was reported in 2006 by Zhao and co-workers [8], this material has fascinated the research community both from the standpoint of fundamental physics and potential device applications [9,10]. In the same year, a first-principle study suggested that ferroelectric properties could be explained without invoking correlated electron physics [11]. In a review, Yang and co-workers [12] highlight the need to understand the relationship between the electronic structure and the magnetic properties, with depth resolution. BFO exhibits high ferroelectric Curie temperature ($T_c$) of 1100 K and a Néel temperature ($T_N$) of 640 K, however, its electric and magnetic properties are highly affected by oxygen vacancies due to valence fluctuations of Fe ions from $Fe^{3+}$ to $Fe^{2+}$, and also volatilization of $Bi^{3+}$ during sintering [13].

$La_{0.7}Sr_{0.3}MnO_3$ (LSMO) is one of the many manganite perovskites studied for their complex electronic and magnetic properties [14,15], with Jahn-Teller distortions playing an important role in the description of the electronic structure of these systems [16]. The material presents half-metallic ferromagnetism with a relatively high $T_C = 370$ K that gives rise to a large negative magnetoresistance [17]. Magnetic and electronic transport properties of LSMO are sensitive to thickness, growth techniques and parameters, showing different types of anisotropy depending on them [18,19].

A BFO/LSMO heterostructure is a prototypical candidate to study so-called hybrid multiferroics, since at the interface there is the coexistence of the ferroelectric behavior of BFO and the ferromagnetic properties of LSMO. The magnetoelectric coupling allows for an electric-field control of magnetism with huge implication in electronics and spintronics applications [20]. One question that usually arises is whether new electronic or magnetic states emerge at the layer-by-layer interfaces between BFO and LSMO, and what is the character of these new states [20]. Interdiffusion across the interfaces also plays a major role in magnetic heterostructures, where rough or sharp interfaces give rise to different magnetic properties [21–23]. The chemical stabilities in BFO and the anisotropy in LSMO properties require chemical characterization probes, such as X-ray photoelectron spectroscopy (XPS), in order to carefully determine the electronic structure of these systems. It is also desirable to extend the characterization within a depth precision of the single unit cell scale, discriminating between signal from the surface, from the bulk, and, in case of heterostructures, from the interfaces. BFO and LSMO heterostructures have been widely characterized and their electronic structure has been studied by conventional spectroscopy techniques and by first principles theoretical approaches [20,21,24–28].

Here, we extend the past experimental results using high-precision depth selectivity across the buried BFO/LSMO interface using standing-wave hard X-ray photoemission spectroscopy (SW-HAXPES) [29–32]. We focus our attention on the same system employing an advanced photoemission spectroscopy technique with high-depth precision, which allows for depth mapping of interfaces. The depth selectivity is achieved by exciting X-ray standing-wave from the first-order Bragg reflection of the BFO/LSMO multilayer superlattice. In combination with higher information depth of hard X-ray photoemission, the technique is capable of probing deep interfaces with sub-nm accuracy.

## 2. Methods

### A. Sample synthesis

BFO/LSMO superlattices were fabricated on $TiO_2$-terminated (001) single-crystal $SrTiO_3$ substrates by pulsed laser deposition, with reflection high-energy electron diffraction (RHEED) control of the growth process, as discussed below and in the Supplemental Material 1. The growth process of $BiFeO_3$ and $La_{0.7}Sr_{0.3}MnO_3$ were optimized in previous studies [24,33] to result in an ideal unit-cell-controlled layer-by-layer growth and bulk-like magnetic and transport properties. The RHEED intensity oscillations during growth of successive layers indicate control on the unit cell (u.c.) scale and a layer-by-layer growth mode. The superlattices consist of 10 periods of a bilayer consisting of 6 u.c BFO (24.2 Å) layer and a 6 u.c. LSMO (23.1 Å) layer. After growth at 670 °C, the heterostructures were slowly cooled to room temperature in 1 bar of oxygen at a rate of 5 K/min. to optimize the oxidation level. The low surface roughness and high crystallinity of the complete superlattice was confirmed by atomic force microscopy and X-ray diffraction, see Supplemental Materials 2 and 3. Clear ferromagnetic behavior can be observed up to ~230 K (see Supplemental Material 4), which is in good agreement with observations for single 6 unit cells thin LSMO layers [33]. Furthermore, clear exchange bias behavior can be observed in good agreement with BFO/LSMO bilayer systems [24].

### B. Standing-wave excited photoemission

The standing-wave experiments rely on the interference between incident and reflected x-rays, with the incidence angle $\theta_x$ being varied around the first-order Bragg condition of the superlattice under study. The relevant Bragg equation is $\lambda_x = 2d_{ML} \sin(\theta_x)$, where $\lambda_x$ is the wavelength of the incident x-ray photon, $d_{ML}$ is the period or bilayer thickness in the multilayer, and $\theta_x$ is the incidence angle relative to the sample surface. Figure 1 shows a sketch of the superlattice with key parameters of the experiment. In this condition, the period of the resulting SW electric-field intensity $|E|^2 \equiv \lambda_{SW}$ is very close to the period $d_{ML}$ of the superlattice. The SW electric-field varies sinusoidally with the sample depth and can be swept through it in order to provide depth resolution to XPS. This can be achieved by either scanning the incidence angle $\theta_x$ (or take-off angle $\theta_{xe}$) or the photon wavelength $\lambda_x$ over the Bragg reflection. Both approaches induce a phase change in the SW electric-field and the anti-nodes are shifted vertically down the sample by one-half of its period. Finally, the SW will enhance or reduce the photoemission signal from different depths following an unique shape of the electric-field. Multiple core-level and valence band photoemission intensity rocking curves (RC) can be generated and the unique phase information contained in them can be translated into a depth distribution of all the elements in the sample. Beyond the Bragg peak modulation, an additional fine structure called Kiessig fringes is generated by interference effects due to the reflection of top and bottom surfaces of the superlattice. In this case, the relevant equation is $m\lambda_x = 2D_{ML}\sin(\theta_{K,m})$, where $m$ is the $m^{th}$-order interference, $D_{ML} = Nd_{ML}$ is the total thickness of the multilayer mirror with N repetitions, and $\theta_{K,m}$ are the incidence angles corresponding to the reflection order.

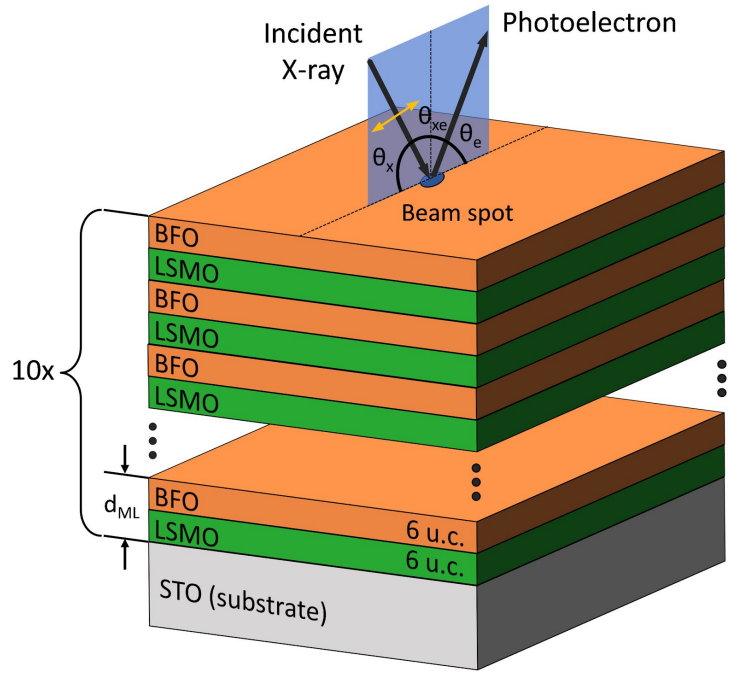

**Figure 1**: The experimental geometry and the structure of the [6 u.c. BFO / 6 u.c LSMO]$_{\times 10}$ sample with various key parameters defined for a standing-wave study using the multilayer as standing-wave generator.

The SW-HAXPES measurements were performed at the undulator beamline BL15XU of the synchrotron radiation facility SPring-8. All spectra were measured using a photon energy of 5953.4 eV, with the total energy resolution at around 240 meV. Energy calibration was done using an Au reference sample and measuring the kinetic energy of the Fermi edge. The beam was focused to a 25 x 25 μm spot, which at

~1° grazing incidence angle translates to 30 x 1800 μm projection on the sample. The radiation polarization was in the photoemission plane with the polarization vector pointing in the direction of the analyzer. Information depth of the photoemission experiments is proportional to electron attenuation length, which in single-scattering approximation translates to the inelastic mean free path (IMFP) [34]. The IMFP of photoelectrons was estimated using the TPP-2M formula and ranges from 50 to 70 Å accounting for large variation in kinetic energies of corresponding core-level photoelectrons [35]. Resulting information depth of the experiments (~95% of the signal) is then covering ca. first three periods of the superlattice. Sample was mounted on a 4-axis manipulator allowing for translation and variation of the polar angle and all measurements were done at room temperature. The photoemitted electrons were captured and analyzed for their kinetic energy by the VG Scienta R4000 spectrometer. The Bragg angle $\theta_B$ for the first-order reflection in the BFO/LSMO sample studied here is around 1.35°, and the standing-wave measurements were done by scanning the incidence angle between 0 and 1.84°.

## 3. Results and discussion

Figure 2(a) shows the Bi $5d_{5/2}$, O 1s, and Sr $2p_{3/2}$ core-level spectra of the superlattice at an off-Bragg incidence angle. However, the full SW measurements are performed from the grazing incidence past the first-order Bragg angle (for this superlattice and hν = 5953.4 eV Bragg angle is 1.35°), spanning from 0° to 1.84° in 0.01° steps. The experimental data are sets of core-level and valence band spectra measured at this interval of incidence angles. Figure 2(b) shows an example of the core-levels measured at three different angles in the vicinity of the Bragg angle. The experimental part of the SW technique then tracks the intensity modulations of the core-level peaks as a function of the incident angle, so-called rocking curves (RCs).

The different components present under one apparent single peak can be hard to discern from each other in a single spectrum, but their particular modulation as a function of incident angle allows for that separation. The spectral deconvolution into low binding energy (LBE) and high binding energy (HBE) components of Bi 5d, O 1s and Sr 2p was performed considering Voigt functions with constrained widths and positions, with more information in Supplemental Material 5. In the Bi 5d spectrum the main LBE component is at 25.81 eV, while the HBE component sits at around 25.97 eV. The energetic separation between the fitted components is 0.16 eV. For O 1s, the main LBE component appears at 529.48 eV, while the HBE is 0.49 eV above that. Finally, the main component for Sr 2p is located at 1939.20 eV, while the high binding energy shoulder is shifted by 0.98 eV. The area under the resulting fitted peaks is then integrated for each x-ray grazing incidence angle, leading to the non-monotonic rocking curves that are shown in Figure 2(c). Here, only the first-order Bragg region is shown. The dashed lines denote the incident angles of the spectra in Figure 2(b). The rocking curves exhibit different modulations for the spectral species, indicating that the contributions originate from different depths of the sample. To understand details of these modulations, we now turn our attention to a full set of X-ray optical and photoemission simulations.

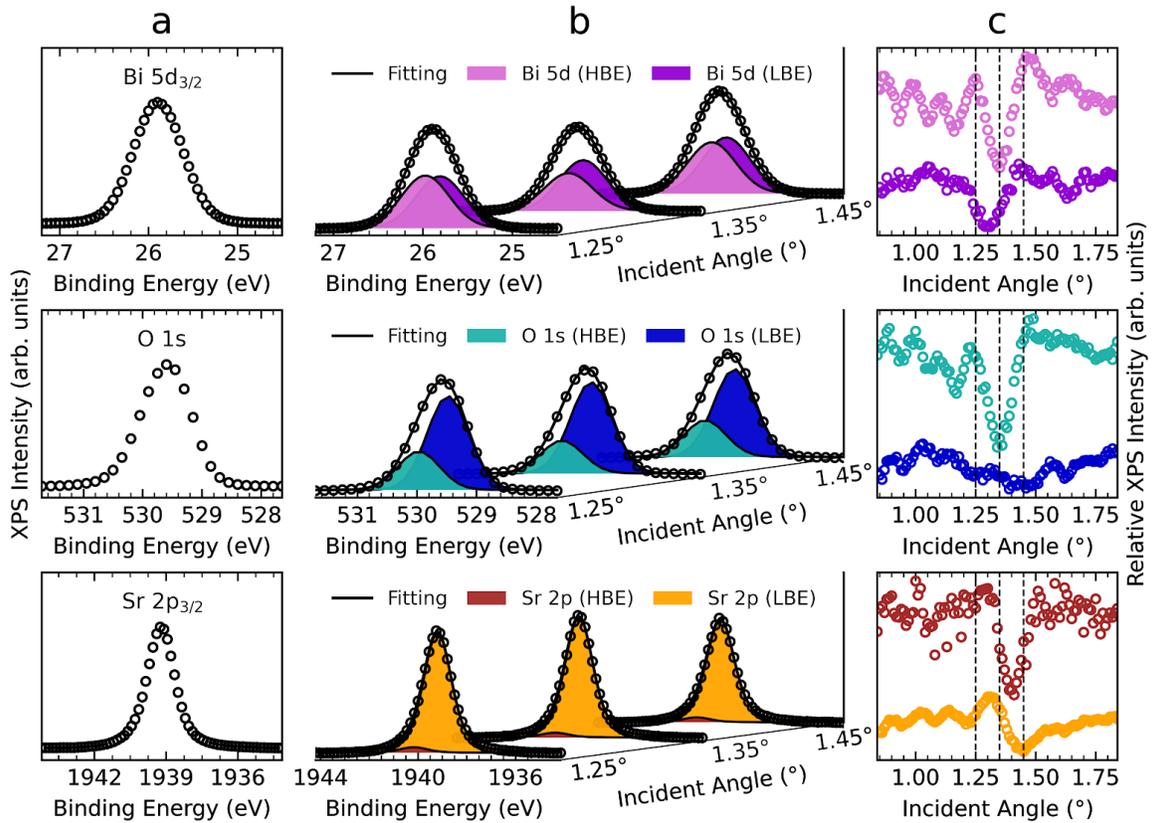

**Figure 2**: Standing-wave photoemission spectroscopy experiment of the [6 u.c. BFO / 6 u.c LSMO]$_{\times 10}$ multilayer measured with a photon energy of 5953.4 eV. (a) Core level spectroscopy of Bi 5d, O 1s, and Sr $2p_{3/2}$ at a specific incident angle (open circles). (b) Core level spectra (open circles) taken at three indicated incident angles and the calculated fitted curves (solid black lines). All spectra present a lower (LBE) and a higher binding energy component (HBE) (filled curves). (c) Rocking curves showing the photoemission signal of each component as a function of the incident angle. The dashed lines indicate the incident angles shown on (b).

In order to determine the depths for each of these species, the optimization of the sample structural model was done in an iterative manner by testing several thousands of choices of sample configurations driven by the so-called black box optimizer algorithm [36,37]. The structural parameters of the model comprise the thicknesses of each of the layers as well as interdiffusion lengths between the individual layers. Using this sample structural model, photoemission rocking curves are then calculated and compared against the experimental rocking curves until the best fit is achieved. Figure 3(a) shows color plots of the standing-wave modulated electric field strength $|E|^2$ within the first three bilayers of the [6 u.c. BFO / 6 u.c LSMO]$_{\times 10}$ multilayer as a function of depth and incidence angle. The angles shown are around the first-order Bragg angle at about 1.35°. Figure 3(b) shows the best-fit sample structural model. Figure 3(c) shows the experimental and calculated core-level rocking curves for all the elements in the multilayer sample. The agreement between the simulation and experimental data is remarkable, in fact both phase and amplitude of the rocking curves, including finer features, such as the Kiessig fringes are very well reproduced.

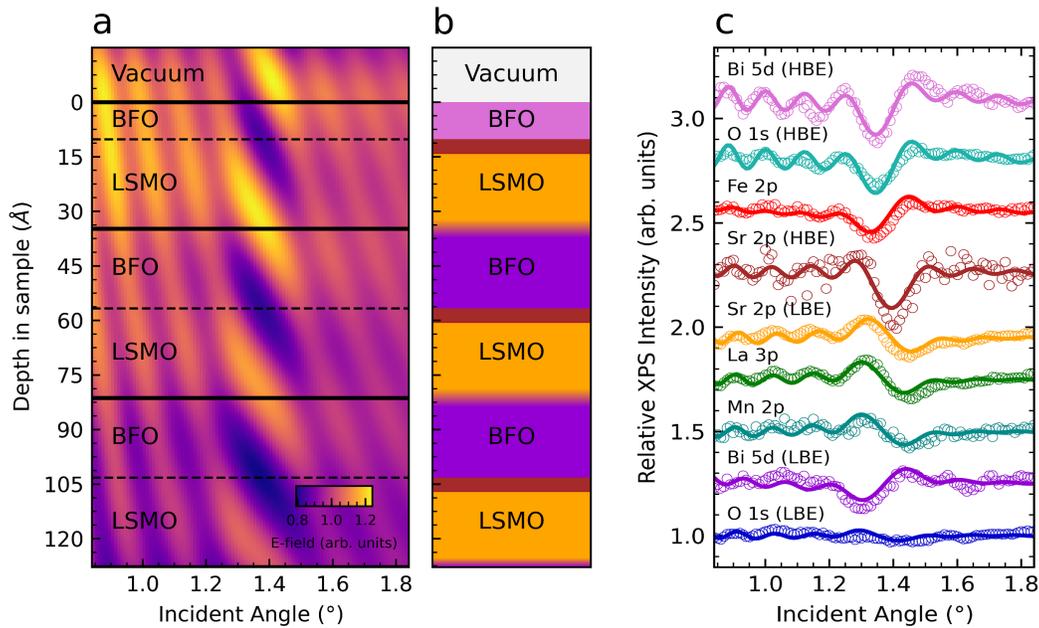

**Figure 3**: Optimization of the sample structure using the standing-wave rocking curves. (a) The calculated standing-wave electric field strength ($|E|^2$) as a function of depth in the sample and of X-ray incident angle. Only the top three BFO/LSMO bilayers are shown. (b) Best-fit optimized structure with its constituents: the top-layer BFO, the BFO/LSMO interface, and the bulk LSMO and BFO. The color gradient at LSMO/BFO interfaces denotes interfacial diffusion. (c) Experimental and calculated rocking curves for various multiple levels assuming the layer structure shown in (b).

There are several observations on the chemical structure of the sample that are yielded by the optimization process of the core-level spectra rocking curves just described. Firstly, we distinguish between the two distinct interfaces in our superlattice: BFO on top of LSMO and LSMO on top of BFO. The interfaces where BFO is on top are sharp and abrupt (to the limit of the SW method accuracy). In contrast, the interfaces with LSMO on top exhibited 1.2 u.c. (4.8 Å) interdiffusion length. Another observation, which will have further implications for our discussion of the valence band data, is that the abrupt BFO/LSMO interface is Sr rich, being populated by the Sr species exhibiting a HBE component in Sr 2p spectra. This chemically distinctive Sr species is contained within 1 u.c. from the interface and one can speculate its role in the electronic coupling between BFO and LSMO. Finally, the top $BiFeO_3$ layer contains exclusively Bi atoms with a HBE component in Bi 5d core-level spectra. This is not surprising, since the sample was transferred through air before measurements and Bi in $BiFeO_3$ could possibly be further oxidized. All buried BFO layers within probing depth of our HAXPES measurements then contained Bi atoms with LBE Bi 5d spectral signature, which is further discussed in Supplemental Material 6, which in detail describes the near total reflection measurements.

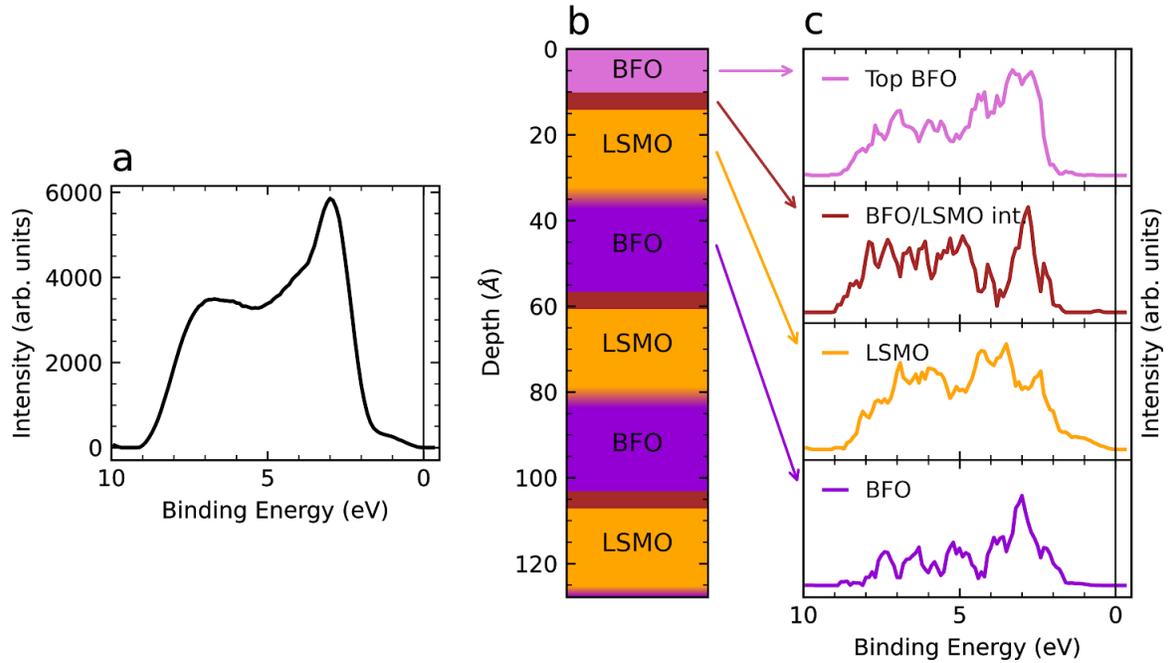

**Figure 4**: Standing-wave valence-band photoemission spectroscopy of the [6 u.c. BFO / 6 u.c LSMO]$_{\times 10}$ multilayer sample. (a) Angle-integrated valence band spectrum. (b) Sketch that indicates the slices used as a deconvolution basis. (c) Slice-by-slice valence band contribution.

We will now proceed to the last step of the standing-wave data analysis, which allows the depth decomposition of valence spectral weight contributions of the individual layers and interfaces. Figure 4(a) shows the total valence band photoemission spectrum of the sample integrated across all incidence and emission angles. This corresponds to matrix element weighted density of states and the vertical line denotes the Fermi energy. One can identify features corresponding to the crystal-field split transition metal states between 0 and 5 eV, while the O 2p-Fe 3d and O 2p-Mn 3d hybridized states appear below 5 eV.

The region of the samples that were used for the decomposition of the valence band spectra are shown in Figure 4(b). The deconvolution procedure specifically targets 4 chemically distinctive regions of the sample: 1. top BFO layer; 2. the 1 u.c. of BFO/LSMO interface; 3. "bulk" LSMO layers; 4. buried BFO layers. The corresponding rocking curves are shown in Fig. 3(c) can be used as base vectors to decompose the valence band into components originating from each respective layer [38]. The results of the decomposition are shown in Figure 4(c).

As expected, both top and buried BiFeO$_3$ layers show a band gap with valence band maximum located at 1.6 eV. This is in agreement with the band gap and location of the Fermi energy reported in literature [39,40]. On the contrary, the La$_{0.7}$Sr$_{0.3}$MnO$_3$ layer is metallic with a clearly distinctive e$_g$ and t$_{2g}$ states of Mn populating Fermi edge and 1.5 eV range below. As one of the results, the decomposition procedure allows for determining band alignment between BFO and LSMO layers and observed band offsets are 1.6 eV in both cases. One could assume a different band offsets for BFO/LSMO and LSMO/BFO interfaces, in case that BFO would be electrostatically polarized out of plane, but our valence band decomposition is not sensitive to such information. Naturally, the top BFO/LSMO has a different band offset when compared to the rest of BFO/LSMO, following the different chemistry of the surface layer. Interestingly, in region 2, which is a single terminating unit cell between BFO and LSMO, an in-gap state near Fermi edge can be observed. As discussed earlier, this region is chemically distinctive by a different HBE species of Sr. One can speculate that this in-gap state is responsible for the electronic coupling between BFO and LSMO and can be explained by a charge transfer across the interface from Fe to Sr atoms. First principle calculations will be used to confirm this hypothesis.

## 4. Conclusions

Standing-wave hard X-ray photoemission was used to investigate the depth-resolved chemical and electronic composition of the hybrid multiferroic system BFO/LSMO and to identify the phenomena at the interface between these two perovskite oxides. Two X-ray optical effects were used to enhance depth selectivity of the photoemission experiments — near total reflection and Bragg reflection — for creating a non-monotonous X-ray strength profile inside the multilayer sample. In terms of roughness, the interfaces between BFO/LSMO and LSMO/BFO are different, with the first named (BFO on top) being sharp and abrupt and the latter (LSMO on top) having interdiffusion length of around 1.2 u.c. (4.8 Å). Apart from differences in roughness, the two interfaces also exhibit different chemical gradients, with the BFO/LSMO interface being Sr-terminated by a spectroscopically distinctive high binding energy component in Sr core-level spectra, which is spatially contained within 1 u.c. from the interface. As expected, due to sample exposure to the ambient atmosphere and no vacuum treatment prior to our experiments, Bi is present in a higher oxidation state in the topmost layer, contrary to other two buried BFO layers that are within the experimental probing depth.

From the electronic point of view, unique valence band features were observed for bulk-BFO, bulk-LSMO and their interfaces. The in-gap state responsible for the electronic coupling between BFO and LSMO is observed at the BFO/LSMO interface and is connected to a charge transfer between LSMO and BFO layers. Apart from information on chemical and gradients in the sample, valence band decomposition based on Bragg-reflection standing-wave measurement also allowed for a direct observation of band alignment between BFO and LSMO layers.

In conclusion, hard X-ray standing-wave excited photoemission experiments were used to obtain layer-resolved chemical and electronic structure of the BFO/LSMO hybrid multi-ferroic system with the depth resolution on the single unit cell level. To our knowledge, standing-wave photoemission is the only non-destructive method capable of probing electronic structure with depth-resolution of this level of accuracy. Our detailed analysis revealed the unique electronic signature at the BFO/LSMO interface connected to electronic coupling between these two ferroics. The equivalent investigations can be successfully applied to a broad class of material such as perovskite complex oxides with emergent interfacial phenomena.

**Acknowledgements**

The SW-HAXPES measurements were performed with the approval of NIMS Synchrotron X-ray Station (Proposal No. 2011A4606). The authors would like to thank the staff of HiSOR, Hiroshima University and JAEA at SPring-8 for the development of HAXPES at BL15XU of SPring-8. H.P.M. has been supported for salary by the U.S. Department of Energy (DOE) under Contract No. DE-SC0014697. S.K. and J.M. would like to thank the CEDAMNF Project financed by the Ministry of Education, Youth and Sports of Czech Republic, Project No. CZ.02.1.01/0.0/0.0/15_003/0000358 and the Czech Science Foundation (GACR), Project No. 20-18725S. A.X.G. acknowledges support from the U.S. Department of Energy, Office of Science, Office of Basic Energy Sciences, Materials Sciences and Engineering Division under award number DE-SC0019297 during the writing of this paper.

**Supplemental Materials**

**1. Superlattice growth**

$BiFeO_3$/$La_{0.7}Sr_{0.3}MnO_3$ (BFO/LSMO) superlattices were fabricated by pulsed laser deposition with reflection high-energy electron diffraction (RHEED) control of the growth process. Atomically smooth $TiO_2$-terminated $SrTiO_3$(100) substrates were prepared by a combined HF-etching/anneal treatment [41]. All substrates had vicinal angles of ~0.1°. Stoichiometric BFO and LSMO targets were ablated at a laser fluence of 1.5 J/cm$^2$ and a repetition rate of 1 Hz. During growth, the substrate was held at 670 °C in an oxygen environment at 2.6 x 10$^{-1}$ mbar. The growth processes were optimized in previous studies [24,33] to result in ideal unit-cell-controlled layer-by-layer growth and bulk-like magnetic/transport properties. After growth, the heterostructures were slowly cooled to room temperature in ~1 bar of oxygen at a rate of 5 K/min to optimize the oxidation level.

**2. Surface topography**

The low level of surface roughness was confirmed by atomic force microscopy (AFM) analysis of the surface of 10 bilayers thick BFO/LSMO superlattice. Figure S1 shows the topography image of a smooth superlattice surface with terraces separated by clear unit cell height steps similar to the surface of the initial $TiO_2$-terminated SrTiO3 (100) substrate.

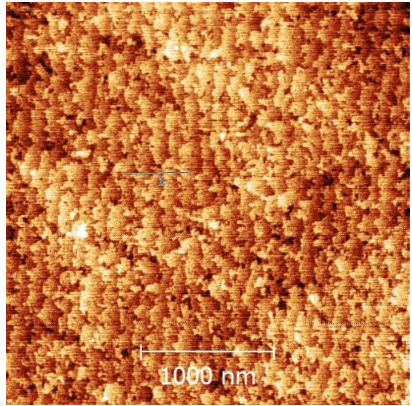

**Figure S1**. Surface topography of a [6 u.c. BFO / 6 u.c. LSMO]$_{\times 10}$ superlattice by AFM of a 3 x 3 μm² area.

## 3. Crystal structure

The epitaxial relation between the individual BFO and LSMO layers in a 10 bilayers thick BFO/LSMO superlattice were studied by X-ray diffraction. The superlattice was in-plane fully strained to the SrTiO$_3$ (100) substrate, as shown by reciprocal space mapping around the (103) SrTiO$_3$ peak (see Figure S2), resulting in an in-plane lattice parameter of 3.905 Å. The out-of-plane superlattice periodicity was determined by a -2 scan around the (002) peak of the SrTiO$_3$ substrate displaying the corresponding superlattice peak SL(002) and the first higher and lower order superlattice diffraction peaks, SL+1 and SL-1. Detailed analysis showed the presence of clear Kiessig fringes alongside the SL(002) superlattice peak indicating a highly ordered crystalline sample with very smooth interfaces and surface, see Figure S2. The superlattice periodicity was determined to 47.3 Å, which agrees very well with 6 unit cells of strained BFO (c-axis 4.04 Å) [24] together with 6 unit cells of strained LSMO (c-axis 3.84 Å) [33].

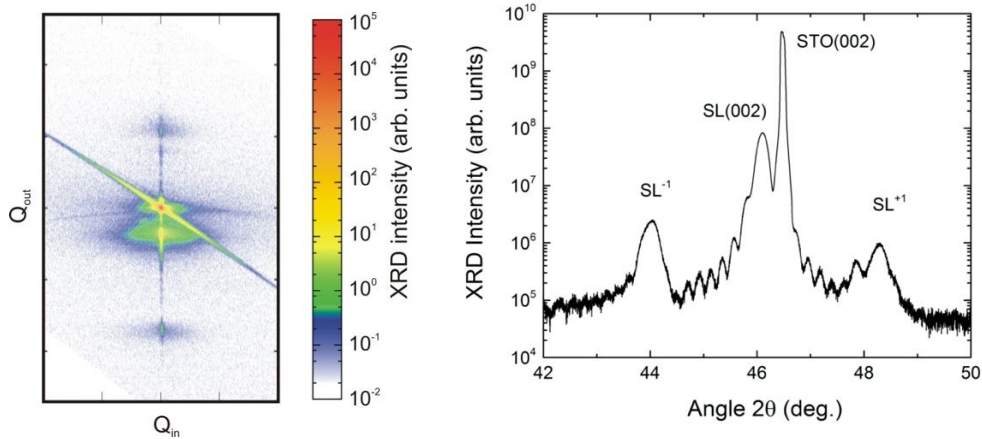

**Figure S2**. X-ray diffraction analysis of a [6 u.c. BFO / 6 u.c. LSMO]$_{\times 10}$ superlattice. (left) Reciprocal space map around the (103) SrTiO$_3$ peak showing a fully in-plane strained superlattice. (right) -2 scan around the (002) SrTiO$_3$ peak showing out-of-plane superlattice (SL) periodicity with Kiessig fringes.

## 4. Ferromagnetism

The magnetic properties of a 10 bilayers thick BFO/LSMO superlattice were measured in a Quantum Design PPMS system. Figure S3 shows typical hysteresis curves at 10 K along the [100] direction after -1 T, +1 T and zero after field cooling from 360 K. Clear ferromagnetic behavior can be observed up to ~230 K in a superlattice consisting of individual LSMO layers of only 6 unit cells, which is in good agreement with observations for single 6 unit cells thin LSMO layers [33]. Furthermore, clear exchange bias behavior can be observed in good agreement with BFO/LSMO bilayer systems [24].

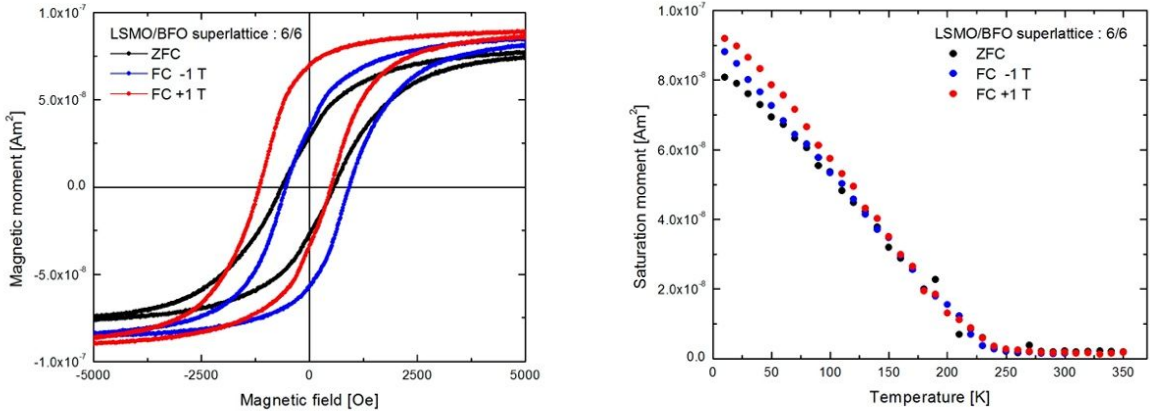

**Figure S3**. Magnetic properties of a [6 u.c. BFO / 6 u.c. LSMO]$_{\times 10}$ superlattice along the [100] direction. (left) Magnetic hysteresis loops at 10 K; (right) temperature dependent saturation magnetization extracted from full hysteresis loops after +1 T, -1 T and zero field cooling from 360 K. The diamagnetic contribution of the SrTiO$_3$ substrate has been subtracted.

## 5. Core level spectroscopy and curve fittings

In order to extract the rocking curves for the intensities of the photoemission standing-wave maps, the core-level spectra of each species in the superlattice are fitted for all incident angles. The procedure to fit the spectra started with the removal of the Shirley background of all measured datasets, which was performed using the KolXPD software. The data was then fitted using a model that consisted of sums of Voigt-function components, parametrized by their binding energy, amplitude, and Gaussian/Lorentzian widths. Each Voigt component is a convolution of Gaussian and Lorentzian profiles, which are required to adequately describe the experimental and core-hole lifetime broadening effects. The fitting was performed using the Python lmfit package.

As discussed previously, due to the standing-wave electric field, the different spectral components in the photoemission spectra can be modulated differently at different depths. To pick up such modulations along the different incident angles, a fine fitting procedure is needed, specially to be able to discern beyond the most apparent components that compose the spectra. One key aspect of the fitting procedure adopted here was that the fitting for each dataset was performed all at once for all incident angles. More than that, the binding energy and broadening parameter values were set to be the same for all incident angles; the amplitudes of the components were left as the only angle-dependent parameters. These constraints enforce that the components are consistent for all incident angles. As an example, Figure S4 shows the experimental spectra measured at 1.35° (open circles) and the corresponding calculated curves, both the total fit (black line) and components (colored lines). The best fitted curves for the Fe $2p_{3/2}$ and Mn $2p_{3/2}$ core-level standing-wave spectra were obtained using four components each, while only two components were necessary for the La $3p_{3/2}$ dataset. For the Bi $5d_{5/2}$, Sr $2p_{3/2}$, and O 1s datasets, two components were used to fit the maps. In these cases, an additional constraint was added so the broadenings are also the same for the two components. The requirement for HBE and LBE components for these core-levels is more evident in the O 1s spectrum, as the HBE shoulder is more apparent. The same was needed for the Bi $5d_{5/2}$ and Sr $2p_{3/2}$ spectra, however, in order to achieve the best fit. The resulting rocking curves were then calculated by integrating and summing all components that composed the spectra, and are shown in the main text.

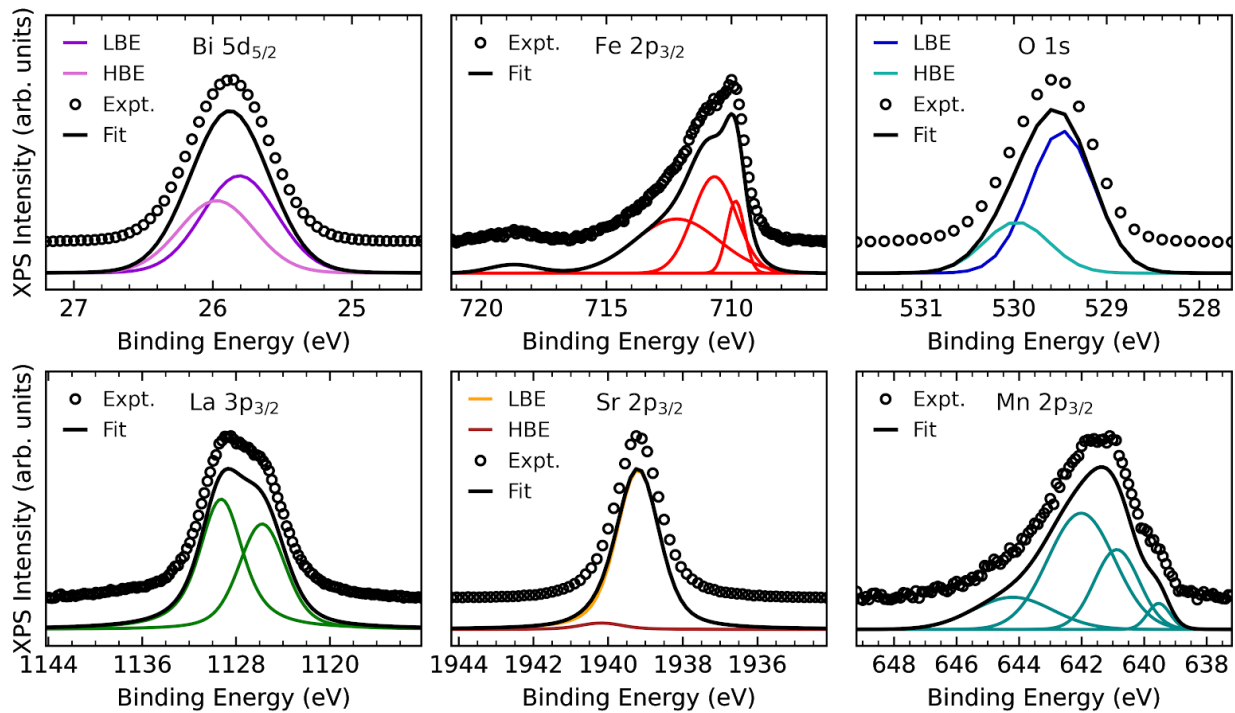

**Figure S4**. Spectra decomposition into Voigt-function components of the core-level photoemission spectra for Bi 5d, Fe 2p, O 1s, La $3p_{3/2}$, Sr $2p_{3/2}$ and Mn $2p_{3/2}$.

## 6. Near total reflection region

Figure S5(a) shows the electric field strength as a function of sample depth and X-ray incidence angle, calculated for the structure reported in the Results section. Here the angle axis spans over the total reflection region including the onset angle. Subsequently, Figure S5(b) shows the photoemission intensities of different core-level electrons near the onset angle. The intensities are shown normalized and correspond to the integrated spectra across all binding energies. The order of the onsets for each core-level follows the depth order of the probed layers. The onset for the Bi 5d and O 1s signals happens first, which shows that the superlattice surface is BiO-rich. After that, the next signal to pick up is Fe 2p. The Sr 2p signal comes next and then the final La 3p and Mn 2p onsets appear last. In accordance with Figure S5(a), the onset at lower angle correlates with lower depth of that particular species. The photoemission onset curves were also separated into low- and high-binding energy components for Bi 5d, O 1s, and Sr 2p. Figure S5(c) and Figure S5(d) show the Bi 5d and O 1s curves with clear different onsets for both components, indicating that the high-binding energy component is originating from the surface layer. Finally, in Figure S5(e) the high-binding energy component of the Sr 2p component is shown to be originated from a lower depth than the main low-binding energy peak.

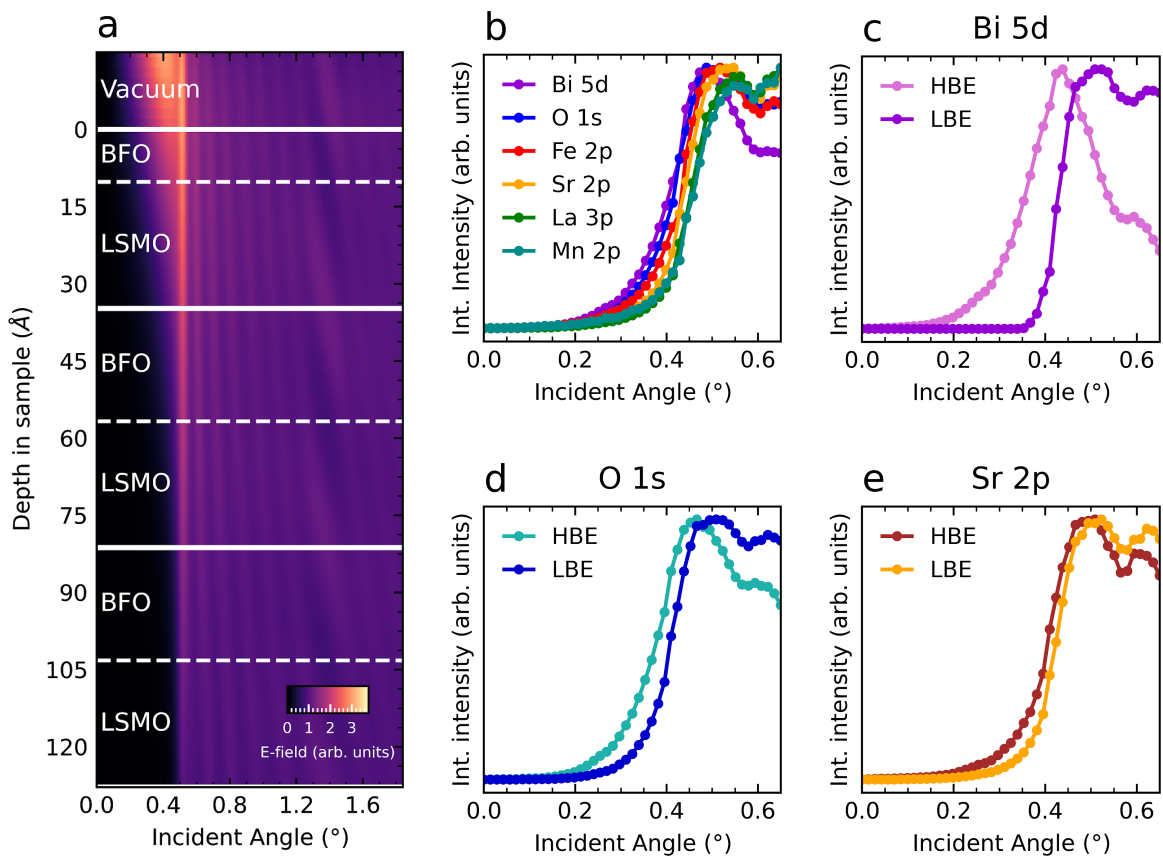

**Figure S5.** Near total reflection region. (a) Calculated electric field strength as a function of sample depth and incidence angle over total reflection region. (b) Photoemission intensities of different core-level electrons near the onset angle. (c-e) Photoemission onset curves for low- and high-binding energy components of (c) Bi 5d, (d) O 1s, and (e) Sr 2p.